\documentclass[a4paper,fleqn,usenatbib]{mnras}
\usepackage{newtxtext,newtxmath}
\usepackage[T1]{fontenc}
\usepackage{ae,aecompl}
\usepackage{graphics}
\usepackage{graphicx}
\usepackage{amsmath}	
\usepackage[flushleft]{threeparttable}
\usepackage{booktabs,caption}
\usepackage{booktabs}
\usepackage{multirow}
\usepackage{siunitx}

\title[4U 1538-522]{Long-term evolution of Cyclotron Line energy in an eclipsing pulsar 4U 1538-522.}

\author[Ruchi et. al.]{
Ruchi Tamang,$^{1}$\thanks{ruchitamang76@gmail.com}
Manoj Ghising,$^{1}$\thanks{E-mail:manojghising26@gmail.com}
Mohammed Tobrej,$^{1}$\thanks{tabrez.md565@gmail.com} 
Binay Rai,$^{1}$\thanks{binayrai21@gmail.com}
\newauthor
Bikash Chandra Paul$^{1}$\thanks{bcpaul@associates.iucaa.in}
\\
$^{1}$Department of Physics, North Bengal University, Siliguri, Darjeeling, WB, 734013, India
\\
}
\pubyear{2021}

\begin{document}
\label{firstpage}
\pagerange{\pageref{firstpage}--\pageref{lastpage}}
\maketitle

\begin{abstract}
 We present the timing and spectral analysis of the HMXB source 4U 1538-522 using NuSTAR observations. One of the observations partially covers the X-ray eclipse of the source along with eclipse ingress. The source is found to spin down at the rate of 0.163 $\pm$ 0.002 s $\text{yr}^{-1}$ between $\sim$ (54973-58603) MJD. It is evident that at time $\sim$ 58620 MJD, a torque reversal occurred, thereafter the source exhibited a spin-up trend at the rate  - (0.305 $\pm$ 0.018) s $\text{yr}^{-1}$ until 59275 MJD. A recent NuSTAR observation finds the pulse period of the source: (526.2341 $\pm$ 0.0041) s. The pulse profile exhibits a transition from double-peaked to single-peaked nature above $\sim$ 30 keV. We analyzed the overall trend of the temporal evolution of fundamental Cyclotron Resonance Scattering Feature (CRSF), $\text{E}_\text{cyc}$, incorporating recent NuSTAR measurements. Initially, during the time span $\sim$ (50452.16-55270.8) MJD, the cyclotron line energy is found to increase at a rate of 0.11 $\pm$ 0.03 keV $\text{yr}^{-1}$ which is further followed by a decrease at a rate - 0.14 $\pm$ 0.01 keV $\text{yr}^{-1}$ between (55270.8-59267) MJD. The combined measurements in the time span (50452.16-59267) MJD reveal that the cyclotron line energy is increasing linearly at a rate of 0.08 $\pm$ 0.02 keV $\text{yr}^{-1}$.
\end{abstract}

% 

%% Select between one and six entries from the list of approved keywords.
%% Don't make up new ones.

%%%%%%%%%%%%%%%%%%%%%%%%%%%%%%%%%%%%%%%%%%%%%%%%%%%
%
%%%%%%%%%%%%%%%%%% BODY OF PAPER %%%%%%%%%%%%%%%%%%
%
\section{Introduction}

In general, High-Mass X-ray Binaries (HMXBs) consist of a compact object orbiting a massive OB class star. The compact object may be either a neutron star (NS) or a black hole. It is a strong X-ray emitter which is a consequence of accretion of matter from the OB companion. HMXBs are further classified into two categories: Be/X-ray binary (those in which the primary is a Be star) and SG/X-ray binary (those in which the primary is a supergiant). In the case of SG/X-ray binary, the X-ray luminosity usually originates due to the strong stellar wind of the optical companion. However, for Be binaries the mass transfer is neither typical wind accretion nor Roche-lobe overflow but is due to a formation of the circumstellar disc fed from material expelled from the rapidly rotating Be star \citep{199}. In the wind-fed system, a persistent luminosity of the  order of $10^{35}-10^{36}\;\text{erg\;s}^{-1}$ is observed as a result of accretion from the stellar wind \citep{32}.

The source 4U 1538-522 is a wind-fed persistent eclipsing X-ray binary source which is known to be formed by a large (17 $M_{\odot}$) B0Iab supergiant \citep{1} and a neutron star. The eclipsing nature of the source was established by \textit{Uhuru} (see, e.g., \cite{33}), while the X-ray pulsations were discovered by \cite{34} and \cite{2}. The spin period of the neutron star is associated with significant variabilities. When the source was discovered, the spin period of the source was estimated to be $ (528.93\pm0.10) $ s \citep{2}, and it was characterized by a spin down feature for more than a decade. The source was observed to undergo torque reversals in 1998 \citep{3} and 2009 \citep{4}. Its orbit is characterized by an inclination of (67 $\pm$ 1) \textdegree \citep{5} and the orbital period of the source is  $(3.75\pm0.15)$ days with an X-ray eclipse lasting for $\sim$ 0.6 days \citep{2, 6}. The orbital shape of the system is not clearly known. Various interpretations have been made regarding the nature of the orbit. Circular nature  was reported by \cite{7}; \cite{8}, while \cite{29}; \cite{6} have reported an eccentricity of 0.17–0.18. \cite{30} have made use of both circular and elliptical orbit parameters for estimating the mass of the neutron star using optical light curves and radial velocity measurements. The source is comprised of supergiant donor which is not overfilling its Roche Lobe \citep{1}. As a result the matter accretion may not occur via  the formation of a persistent disk. The phenomenon is similar to other wind-fed systems viz., Vela X-1 and 4U 1907 + 09 that exhibit short-time variability in terms of dips and flaring activities. The persistent X-ray luminosity of the source was estimated to be $\sim 2\times 10^{36}\;\text{erg\;s}^{-1}$ assuming a distance of $\sim$ 6.4 kpc  by \cite{2}. In our study, we have considered the distance of the source as ($6.82\pm0.02$) kpc as reported by \textit{Gaia} parallax measurements {\citep{1222}}.

 Cyclotron Resonance Scattering Feature (CRSF) is a characteristic absorption feature found to originate in the magnetic poles of the neutron star. The estimated value of the centroid energy of CRSF can be used to determine the magnetic field strength on the surface of the neutron star using the relation,                    

  $\text{B}=\dfrac{\text{E}_\text{cyc}(1+\text{z})}{11.57}\;\times\;10^{12}$ G, \;\;\;\;\;\;\;\;\;\;\;\;\;\;\;\;\;\;\;\;\;\;\;\;\;\; (1)
                       
where $\text{E}_\text{cyc}$ is the cyclotron line energy in the unit of keV \citep{119} and $\text{z}$ is the Gravitational red shift parameter. The value of $\text{z}$ is $\sim$ 0.15 for a neutron star with estimated mass 1.0 $\pm$ 0.10 $\text{M}_{\odot}$ \citep{30} and having a radius of 10 km.

The Cyclotron Resonance Scattering Feature at $\sim20$ keV  was first discovered by \cite{21} in the source by \textit{Ginga} observation. An additional absorption feature at $\sim$51 keV was identified by \cite{22} using \textit{BeppoSAX} observations which was later confirmed by \cite{31} as the first harmonic of the CRSF. The results of the phase-resolved analysis corresponding to the fundamental CRSF was first reported by \cite{23} using \textit{Suzaku} observations. \cite{24} reported a linear increment in the CRSF energy by $\sim$1.5 keV for the observations between 1996 and 2012. ASTROSAT measurements by \cite{18} were also consistent with the results obtained by \cite{24}. Thus, the study of evolution of the cyclotron line energy is an interesting matter of study.

In this paper, we present the detailed spectral and timing analysis of the source 4U 1538-522 using NuSTAR observations. The observation and data reduction formalism are discussed in brief in section 2. Timing analysis, spectral analysis, spectral changes in eclipse phase of the source are presented in sections 3, 4, and 5 respectively. Finally, in section 6 we discuss the results obtained.

\section{Observations and Data Reduction}

The most recent observation of the source 4U 1538-522 by \textit{Nuclear Spectroscopic Telescope Array} (NuSTAR) was done on February 16, 2021 and February 22, 2021. We have considered four NuSTAR observations out of which two recent NuSTAR observations are discussed elaborately. Further, we have also considered publicly available  \textit{Fermi}/GBM data for understanding the evolution of spin period of the source.

\subsection{NUSTAR}

The \textit{Nuclear Spectroscopic Telescope Array} (NuSTAR), launched in June, 2012 is the first high-energy X-ray focusing telescope that operates in the energy range of (3-79) keV. It is an observatory that consists of two co-aligned grazing-incidence X-ray telescopes. Each telescope has its own focal plane module, named as Focal Plane Module A (\textsc{fpma}) and Focal Plane Module B (\textsc{fpmb}), consisting of a solid state CdZnTe pixelated detector \citep{10}.

   We reduced NuSTAR data following the standard procedure using the NuSTAR Data Analysis Software (NuSTARDAS) of \textsc{heasoft} v6.29\footnote{\url{https://heasarc.gsfc.nasa.gov/docs/software/heasoft/download.html}} and \textsc{caldb} v 20220525. The required clean event files were filtered from an unfiltered event files using the mission specific \textsc{nupipeline} script. These clean event files are used to obtain the source and background images using the ftool \textsc{xselect} and an astronomical imaging and data visualization application \textsc{ds9}\footnote{\url{https://sites.google.com/cfa.harvard.edu/saoimageds9}}. A circular region with the source as the center and of 100 arcsec radius is selected as a source region. In the same manner, another circular region of the same radius but away from the source is considered as a background region. These source and background files are used to extract the required light curves, spectra, \textsc{fits} file, and \textsc{pha} file imposing \textsc{nuproducts} with a binning of 0.01 s. The backgrounds for both \textsc{fpma} and \textsc{fpmb} data are individually subtracted and their light curves are combined using ftool \textsc{lcmath}. 
   
   We consider four available NuSTAR observations coined as 30401028002, 30401025002, 30602024002, and 30602024004 for the analysis of the source and are referred to as Obs I, Obs II, Obs III and Obs IV respectively. The respective date of observation (MJD range), exposure, estimated pulse period along with an orbital phase covered by each observation are presented in Table \ref{1}.
   
   For orbit correction, the NuSTAR light curves are barycentered to the solar system barycenter using ftool \textsc{barycorr}. We fitted the spectra using \textsc{xspec} v12.12.0 \citep{13}.

\begin{table*}
\begin{center}
\begin{tabular}{cllllc}
\hline		
Observatory 	&	Observation ID	&	MJD range	&	Exposure (ks)	&	Pulse Period(s)		&	Orbital phase covered	\\
\hline												
NuSTAR	& 30401028002     &       57611.81-57612.79     &       43.8    &       526.5638        $\pm$0.0052     &       0.823-1.086    \\
	& 30401025002	&	58605.95-58606.84	&	36.9	&	526.7638	$\pm$0.0011	&	0.462-0.702	\\
	&	30602024002	&	59261.33-59261.84	&	21.8	&	526.3135	$\pm$0.0032	&	0.244-0.381	\\
	&	30602024004	&	59267.04-59267.58	&	21.6	&	526.2341	$\pm$0.0041	&	0.776-0.921	\\

\hline
\end{tabular}
\caption{NuSTAR observations are indicated by their observation IDs, date of observation (in MJD), exposure, their estimated pulse period with an orbital phase covered by each observation. \label{1}}
\end{center}
\end{table*}

\subsection{Fermi/GBM}
 
 The \textit{Fermi} Gamma-Ray Space Telescope was launched in the year 2008 for the study of the gamma-ray sources. It consists of two instruments: Gamma-ray Burst Monitor (GBM) \citep{14} and Large Area Telescope (LAT) \citep{1804}. GBM operates efficiently in an energy range from 8 MeV to 40 MeV and LAT is sensitive in the energy range from 20 MeV to about 300 GeV. We are considering the spin frequency provided by \textit{Fermi}/GBM Accreting Pulsars Program (GAPP) to study the spin frequency evolution of the source.

\section{TIMING ANALYSIS}

 The light curves of the source are extracted in the energy range (3-79) keV by NuSTAR using ftool \textsc{lcurve} with a binning of 0.01 s. The extracted light curves of the NuSTAR observations are considered for determining the pulse period of the source. The Power Density Spectrum (PDS) is generated from the light curve for roughly estimating the pulse period by Fast Fourier Transform (FFT) using command \textsc{powspec}. We obtained the accurate pulse period with the help of the approximated period by implementing the epoch-folding technique \citep{15, 16} using the tool \textsc{efsearch}. This method is based on $\chi^{2}$ maximization technique. Thus, we obtained the best pulse period at $ 526.5638 \pm 0.0052$ s, $ 526.7638 \pm 0.0011$ s, $ 526.3135 \pm 0.0032$ s, and $ 526.2341 \pm 0.0041$ s for Obs I, Obs II, Obs III, and Obs IV respectively which is found to be consistent with the \textit{Fermi}/GBM data. We determined the uncertainty in the measurement of the pulsations by implementing the method described in \cite{17}. For this, we generated 1000 simulated light curves and determined their respective pulsations for each light curve. Hence, the precise pulse period of the source is obtained after computing the standard deviation and standard error of the pulsations obtained. The tool \textsc{efsearch} helps in obtaining the best period but does not show the folded pulse profile. Therefore, we used the tool \textsc{efold} to obtain the folded pulse profile. The pulse profile is normalized about the average count for representative consideration taking the zero-point i.e., phase 0 at the flux minimum to study the comparative variation of the pulse profile with energy and time. In this manner, we generated the folded pulse profile in the energy range of (3-79) keV. To study the energy dependence of the pulse profile we  resolved the (3-79) keV energy range into several energy bands as (3-7) keV, (7-12) keV, (12-18) keV, (18-24) keV, (24-30) keV, (30-50) keV and (50-79) keV. The evolution of pulse profile with energy for NuSTAR Obs III can be clearly seen in Figure \ref{1}. The pulse profile exhibits a double-peaked structure consisting of a primary and secondary peak. The primary peak lying between phase (0-0.6) is observed to become narrower with an increase in energy. The strength of the secondary peak lying between phases (0.6-1.0) decreases with an increase in energy. The pulse profile evolution with energy is consistent as that discussed in \cite{18}.
  
    To study the evolution of pulse profile with time, we generated the pulse profiles in (3-79) keV energy range by folding the background-subtracted light curves at corresponding estimated spin periods for all four NuSTAR observations as shown in Figure \ref{2}. From the figure, we observe only a slight variation in the morphology of the pulse profile with change in luminosity of the source. The small variation in the pulse profile observed with the luminosity indicate that the pulse profile variation with energy for all observations follow similar pattern as shown for Obs III in Figure \ref{1}.
    
    The Pulse Fraction (P.F) is defined as the ratio of the difference to the sum of maximum intensity $\text{P}_\text{max}$ and minimum intensity $\text{P}_\text{min}$ of the pulse profile i.e, P.F = $\frac{\text{P}_\text{max}-\text{P}_\text{min}}{\text{P}_\text{max}+\text{P}_\text{min}}$. The variation of P.F with energy for all NuSTAR observations under consideration is shown in Figure \ref{3}. It is evident from the figure that there is an overall increase in the pulse fraction with increase in an energy consisting of a local minimum near $\sim$ 20 keV. The fall in pulse fraction in the form of a local minima near $\sim$ 20 keV adds to an evidence of the presence of known CRSF feature at $\sim$ 21 keV.
    
    Considering the \textit{Fermi}/GBM data covering $\sim(54973-59275)$ MJD, we plot the time-variation of spin period data as shown in Figure \ref{4}. To roughly estimate the corresponding spin-down rate during $\sim$ (54973-58603) MJD and the spin-up rate during $\sim$ (58620-59275) MJD, we linearly fit the \textit{Fermi}/GBM data for respective time intervals. The source is found to exhibit spin-down trend at the rate of $\dot{\text{P}} $= 0.163 $\pm$ 0.002 s $\text{yr}^{-1}$ until $\sim$ 58603 MJD. On $\sim$ 58620 MJD, there is a torque reversal followed by the spinning up of the source at the rate of $\dot{\text{P}} $= - 0.305 $\pm$ 0.018 s $\text{yr}^{-1}$ until 59275 MJD.
    
 Motivated by the significant energy dependence of the pulse profiles, we examined the Hardness Ratio (HR) with respect to the pulse phase of the source by considering the ratio of the hard energy band (H) to that of the soft energy band (S) i.e., (H/S). As the softer peak practically disappears above 30 keV, in contrast to the dual peaked structure at lower energies, we estimated the hardness ratio (HR) of $\dfrac{(30-50)\;\text{keV}}{(3-30)\;\text{keV}}$ corresponding to all four NuSTAR observations as illustrated in Figure \ref{5}. The HR variation shows that the system becomes harder when the source flux is high and becomes softer when the source flux is low.

\begin{figure}
\begin{center}
\includegraphics[angle=0,scale=0.3]{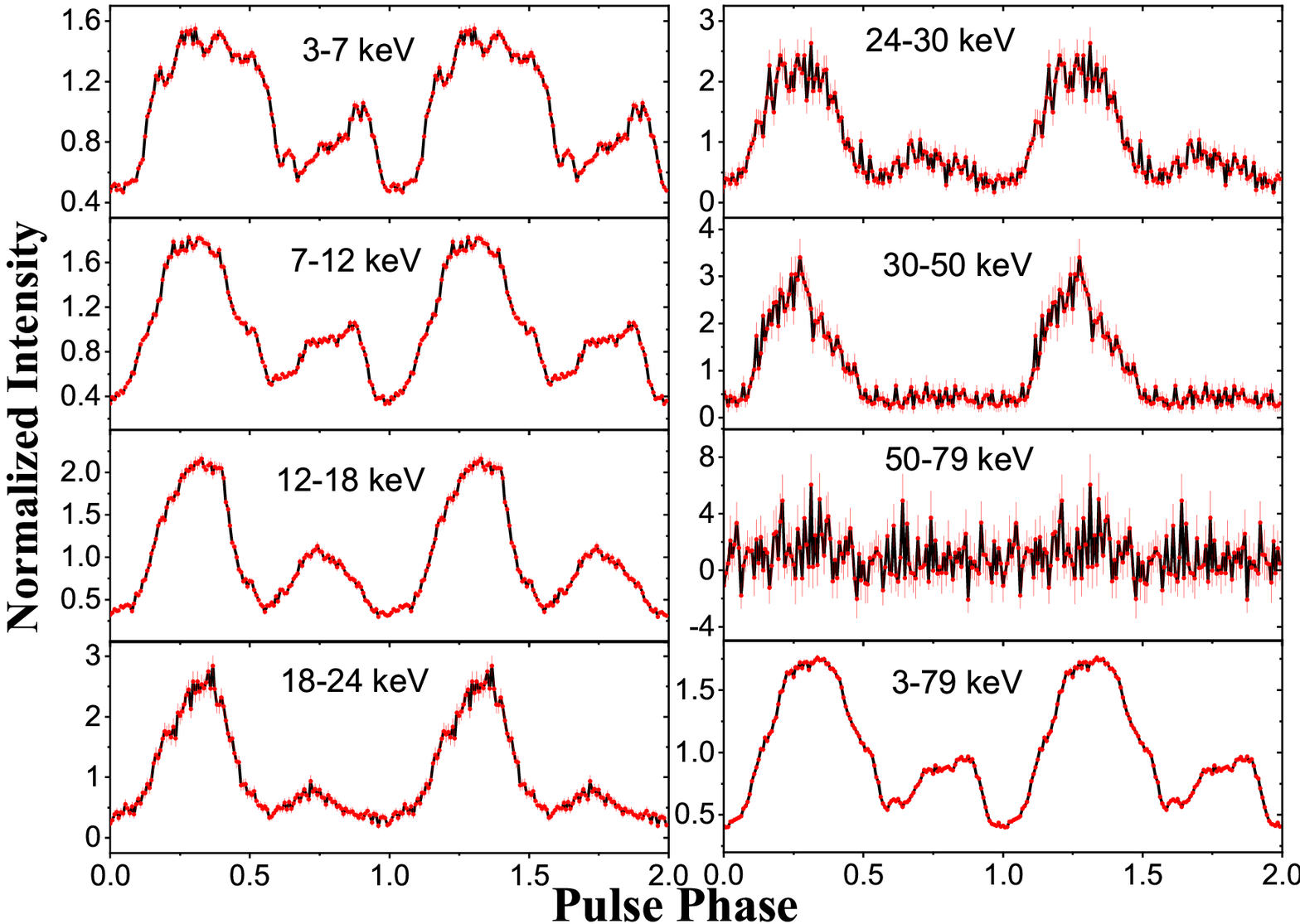}
\caption{Energy resolved pulse profile for NuSTAR Obs III (30602024002) of the source 4U 1538-522. \label{1}}
\end{center}
\end{figure}

\begin{figure}
\begin{center}
\includegraphics[angle=0,scale=0.35]{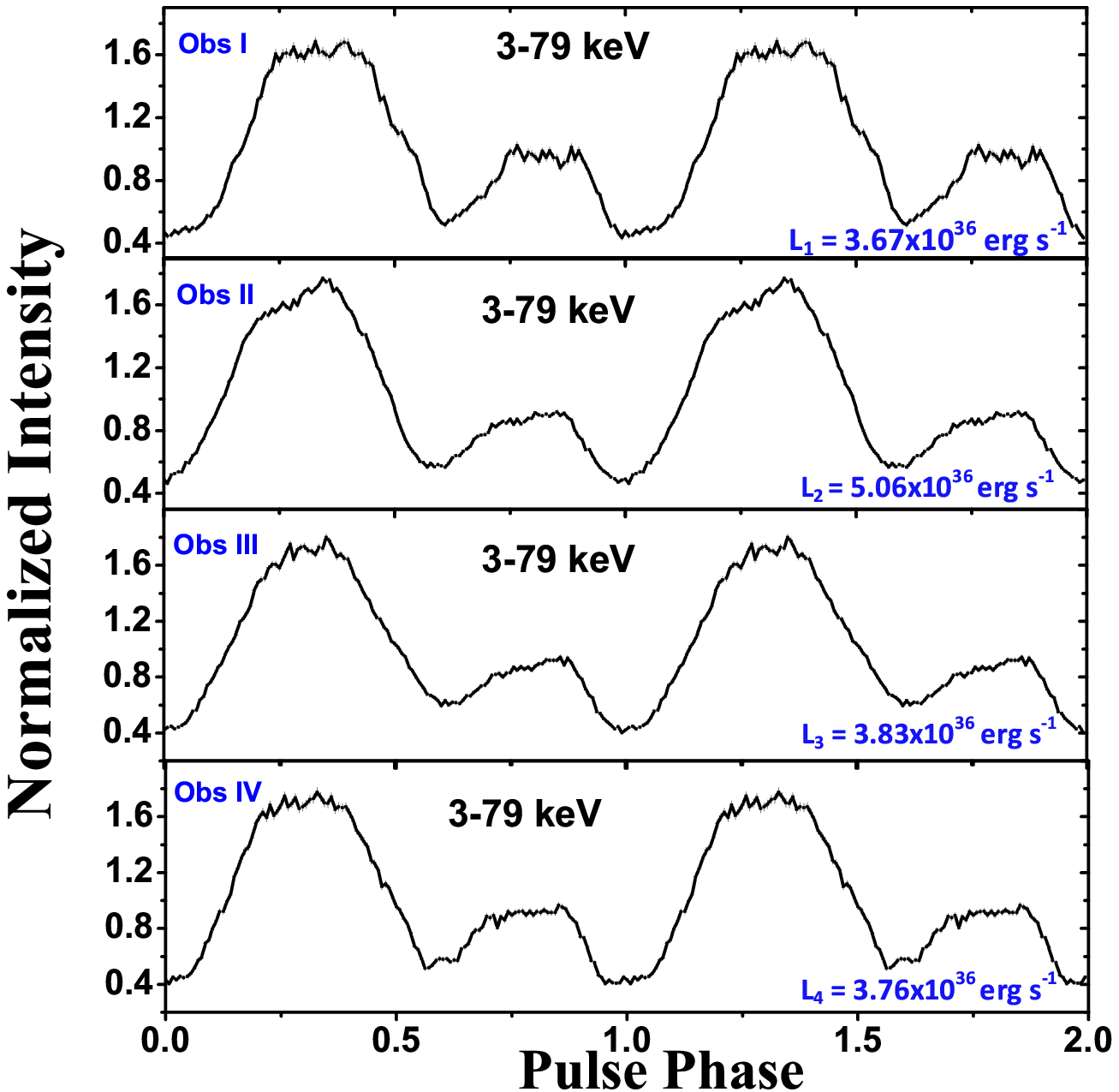}
\caption{Pulse profiles for 4U 1538-522 obtained from the background subtracted light curves of four NuSTAR observations under consideration. \label{2}}
\end{center}
\end{figure}

\begin{figure}
\includegraphics[angle=0,scale=0.32]{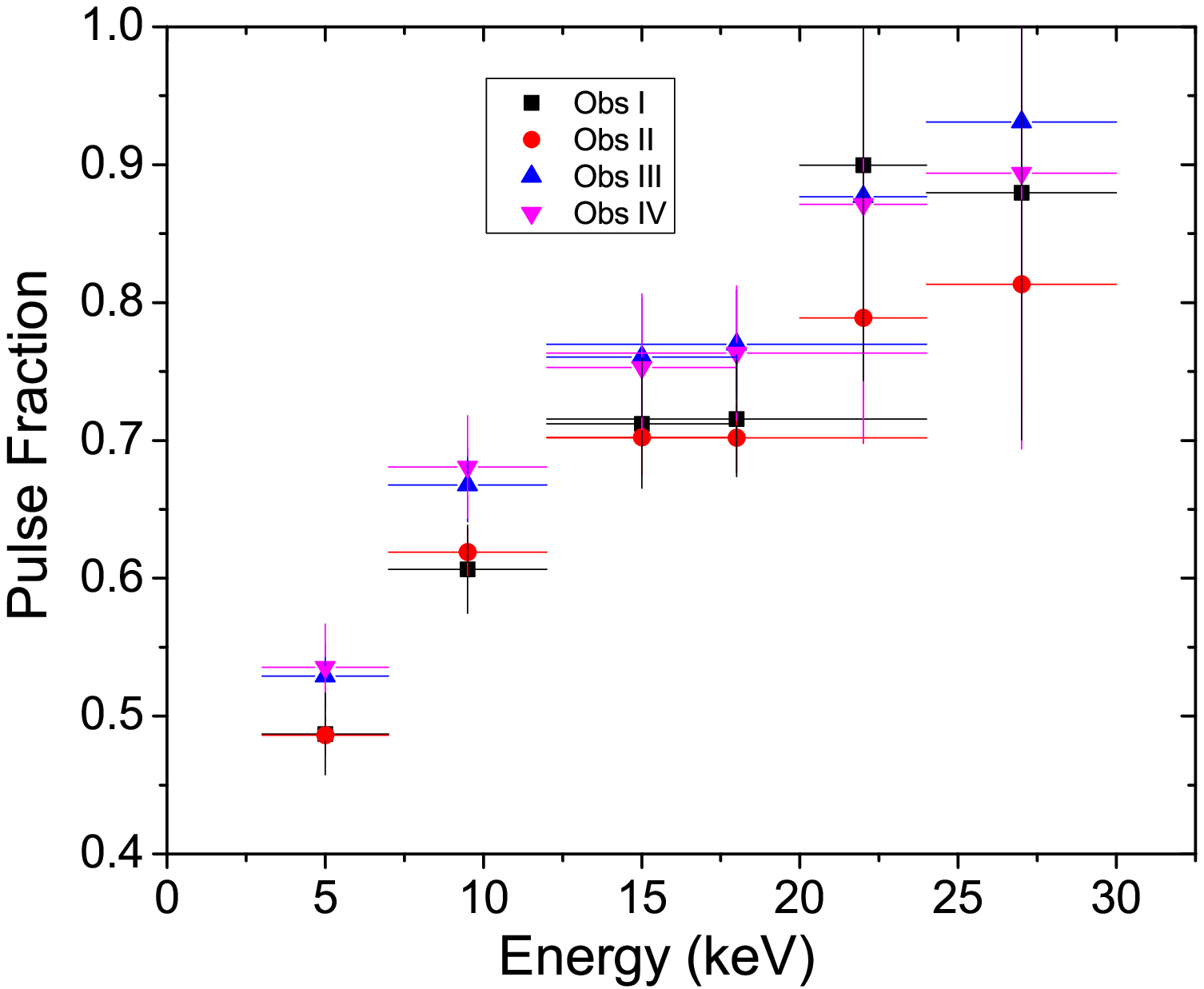}
\caption{Pulse Fraction of the source with respect to the energy (keV) corresponding to four NuSTAR observations. The horizontal error bars on data points represent the energy ranges for which pulse fractions are estimated.The errors quoted are within 3$\sigma$ level uncertainty. \label{3}}
\end{figure}

\begin{figure}
\begin{center}
\includegraphics[angle=0,scale=0.32]{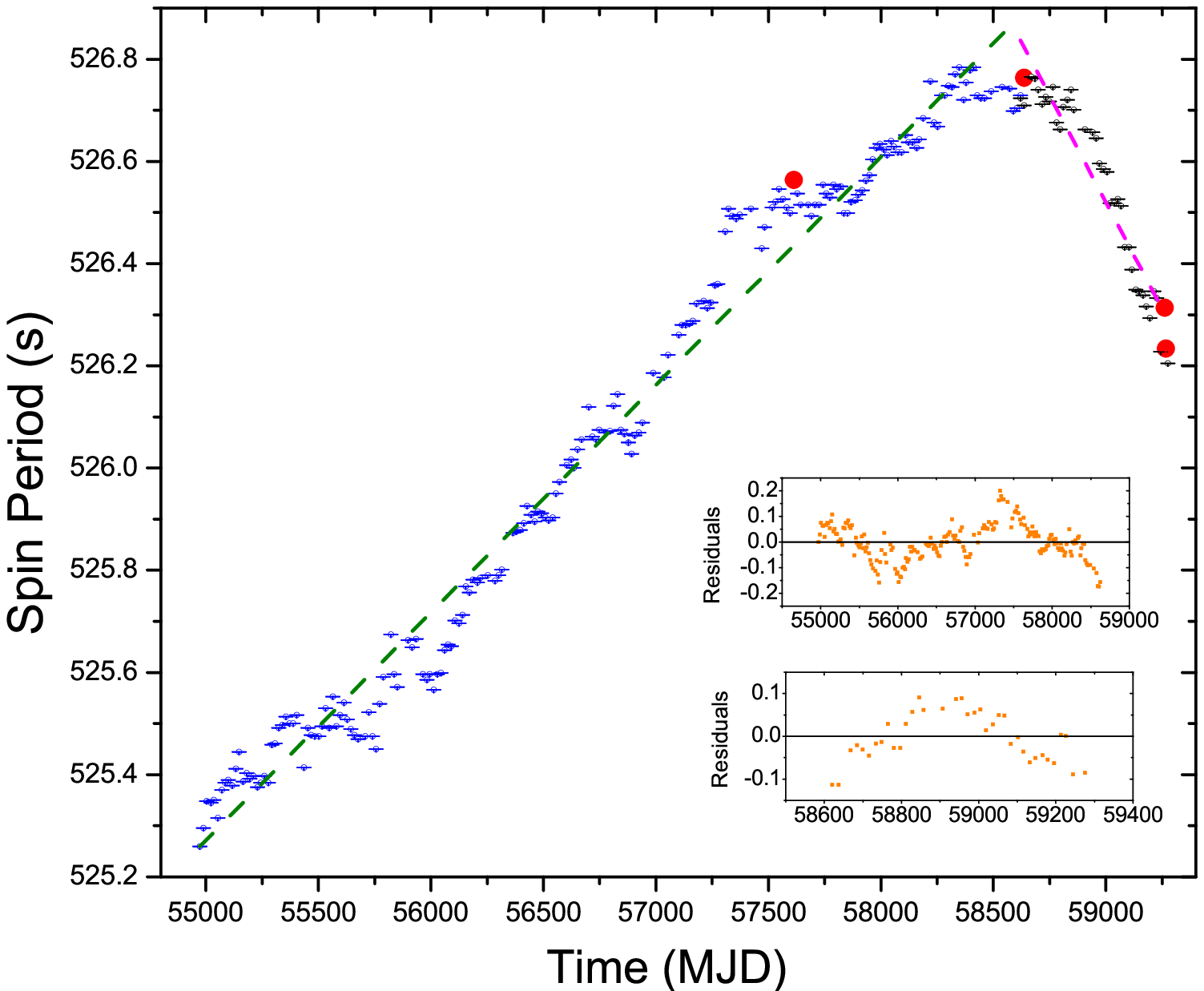}
\caption {Fermi/GBM data showing the temporal variation of spin period of the source. Blue data points represent a spinning down of the source between (54973-58603) MJD after which it undergoes a torque reversal followed by the spinning up of the source between (58620-59275) MJD as denoted by black unfilled circles. An uncertainty associated with spin period is of the order $10^{-8}$ and the corresponding error bars are quoted for each data points. The data points represented by red filled circles corresponds to the four NuSTAR observations. The inset figure shows the residuals for respective fittings. \label{4}}
\end{center}
\end{figure}

\begin{figure}
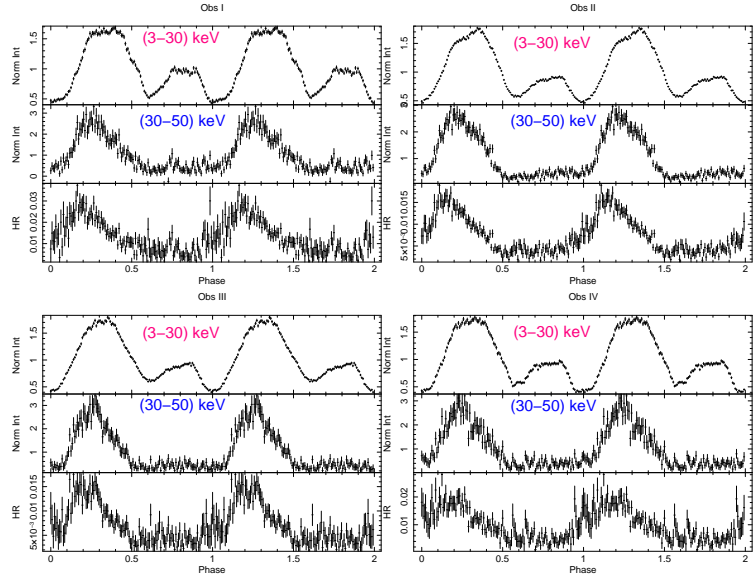

\begin{minipage}{0.24\textwidth}
\includegraphics[height=1.2\columnwidth, angle=270]{H8002}
\end{minipage}
\hspace{0.06\linewidth}
\begin{minipage}{0.24\textwidth}
\includegraphics[height=1.2\columnwidth, angle=270]{h5002}
\end{minipage}
\hspace{0.06\linewidth}
\begin{minipage}{0.24\textwidth}
\includegraphics[height=1.2\columnwidth, angle=270]{h4002}
\end{minipage}
\hspace{0.06\linewidth}
\begin{minipage}{0.24\textwidth}
\includegraphics[height=1.2\columnwidth, angle=270]{h4004}
\end{minipage}
\caption{Variation of hardness ratio with pulse phase. \label{5}}
\end{figure}

\section{SPECTRAL ANALYSIS}
 
\begin{figure*}
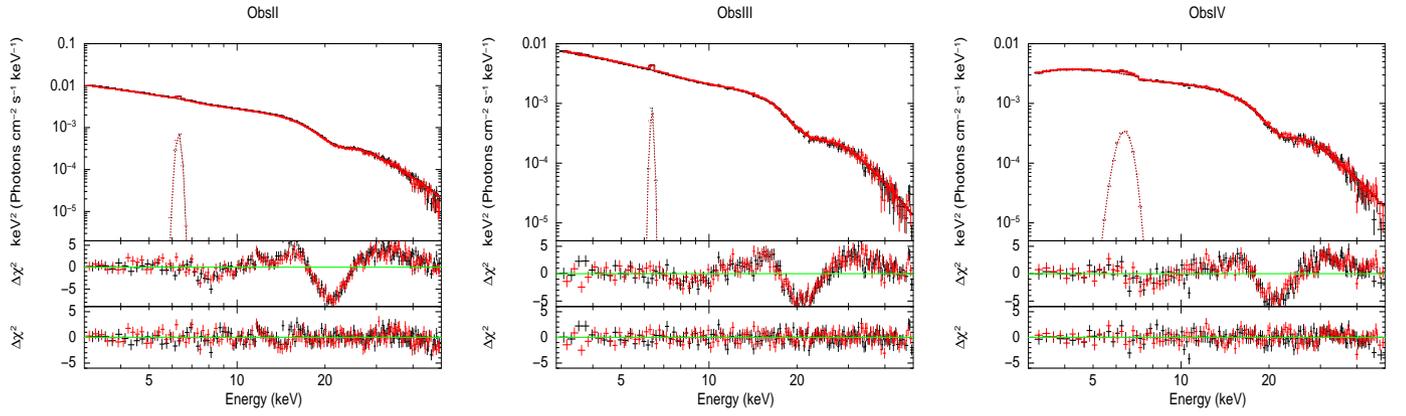

\begin{minipage}{0.3\textwidth}
\includegraphics[width=1.0\linewidth, height=0.25\textheight, angle=270]{5002spec}
\end{minipage}
\hspace{0.04 \linewidth}
\begin{minipage}{0.3\textwidth}
\includegraphics[width=1.0\linewidth, height=0.25\textheight, angle=270]{4002spec}
\end{minipage}
\hspace{0.04\linewidth}
\begin{minipage}{0.3\textwidth}
\includegraphics[width=1.0\linewidth, height=0.25\textheight, angle=270]{4004spec}
\end{minipage}
\caption {An unfolded spectrum of 4U 1538-522 and its approximation with a single continuum model \textsc{constant * phabs * (highecut* powerlaw + gaussian)* gabs* gabs} for Obs II, III, and IV respectively from \textit{left} to \textit{right}. The spectra corresponding to Obs IV is the pre-eclipse spectra. The middle and bottom panels of the figure represent the residuals obtained without and with  \textsc{gabs + gabs} (smooth) model respectively. Red and black indicate the NuSTAR \textsc{fpma} \& \textsc{fpmb} spectra respectively. \label{6}}
\end{figure*}

 The spectral analysis of the source 4U 1538-522 has been carried out for NuSTAR observations II, III and IV. Here we discuss the fitting procedure followed for NuSTAR Obs III (30602024002) which is same for Obs II and Obs IV (pre-eclipse). The NuSTAR (\textsc{fpma} \& \textsc{fpmb}) X-ray spectra are fitted in the (3-50) keV energy band due to background domination above 50 keV. We analyzed spectral binning here considering the optimal binning strategy introduced by \cite{143}. A \textsc{constant} model has been used in spectral fitting in order to account for the instrumental uncertainty. The relative cross-normalization factors between \textsc{fpma} \& \textsc{fpmb} are established by fixing the constant factor for module \textsc{fpma} as unity and allowing the same to vary freely for \textsc{fpmb}. We observed an uncertainty of (1-2) $\%$ which is in good agreement with \cite{19}. Initially, we tried fitting the spectra using the single continuum model \textsc{constant * phabs * (highecut * powerlaw)}. The model combination showed an emission like feature in the (6-7) keV energy range which has been fitted by including the \textsc{gaussian} model. The negative residuals at $\sim$ 21.37 keV reveal a Cyclotron Resonance Scattering Feature (CRSF) and is fitted by adding the Gaussian absorption model (\textsc{gabs}). To smoothen out the residuals we added one more spectral component \textsc{gabs (smooth)}. Hence, we fitted the spectra by using a single continuum model  \textsc{constant * phabs * (highecut* powerlaw + gaussian)* gabs*gabs}. The model estimates the cut-off energy at $\sim$ 16.62 keV and the photon index at $\sim$ 1.09. The emission feature at $\sim$ 6.37 keV is fitted with the \textsc{gaussian} model. The fit statistics ( $\chi^{2}$ per degree of freedom ) is estimated as 1.02. The estimated flux in (3-79) keV energy range is $ (7.05 \pm 0.02) \times 10^{-10}  \text{erg\;cm}^{-2}\;\text{s}^{-1}$   and the corresponding luminosity is $3.92 \times 10^{36} \text{erg\;s}^{-1}$ assuming the source to be at a distance of $\sim 6.82$ kpc. The spectral fitting for Obs II, III and IV (pre-eclipse) are shown in Figure \ref{6}. The corresponding statistical data are presented in Table \ref{2}. In our study, we did not observe the previously confirmed first harmonic of CRSF by \cite{31} at $\sim$ 51 keV.

\begin{table*}
\begin{center}
\resizebox{\linewidth}{!}{%
\begin{tabular}{cllllc}
\hline									
{\bfseries Spectral Parameters}	&	{\bfseries Obs II}		&	{\bfseries Obs III}		&			&	{\bfseries Obs IV} 		&			\\
							\cmidrule{4-6}								
	&			&			&	{\bfseries Pre-eclipse}		&	{\bfseries Ingress}		&	{\bfseries Eclipse}		\\
\hline																
{\bfseries Flux ($10^{-10}\;\text{erg\;cm}^{-2}\;\text{s}^{-1}$)} 	&	9.34	$\pm$0.02	&	7.05	$\pm$0.02	&	6.91	$\pm$0.04	&	4.99	$\pm$0.65	&	0.21	$\pm$0.05	\\
{\bfseries $\text{C}_\text{\textsc{fpma}}$} 	&	1	(fixed)	&	1	(fixed)	&	1	(fixed)	&	1	(fixed)	&	1	(fixed)	\\
{\bfseries $\text{C}_\text{\textsc{fpmb}}$} 	&	0.975	$\pm$0.002	&	1.012	$\pm$0.003	&	1.006	$\pm$0.004	&	1.003	$\pm$0.019	&	1.097	$\pm$0.046	\\
\hline																
{\bfseries nH ($10^{22}\;\text{cm}^{-2}$)}	&	1.02	$\pm$0.12	&	0.49	$\pm$0.22	&	6.19	$\pm$0.51	&	(6.19)		&	(6.19)		\\
{\bfseries $\text{n}_{\text{H},\text{c}}$	($10^{22}\;\text{cm}^{-2}$)} &	--	--	&	--	--	&	--	--	&	70.86	$\pm$6.96	&	140.99	$\pm$80.46	\\
{\bfseries $\text{f}_{\text{p},\text{c}}$}	&	--	--	&	--	--	&	--	--	&	0.94	$\pm$0.01	&	0.69	$\pm$0.1	\\
{\bfseries $\text{E}_\text{cut}$ (keV)}	&	14.04	$\pm$0.09	&	16.62	$\pm$0.21	&	13.61	$\pm$0.13	&	13.77	$\pm$0.89	&	8.03	$\pm$1.2	\\
{\bfseries $\text{E}_\text{fold}$ (keV)}	&	11.95	$\pm$0.09	&	9.75	$\pm$0.35	&	11.04	$\pm$0.20	&	12.98	$\pm$3.37	&	8.13	$\pm$2.44	\\
{\bfseries Photon Index}	&	1.186	$\pm$0.008	&	1.09	$\pm$0.03	&	0.98	$\pm$0.03	&	1.15	$\pm$0.22	&	0.92	$\pm$0.26	\\
{\bfseries $\text{E}_\text{Fe}$(keV)}	&	6.31	$\pm$0.02	&	6.37	$\pm$0.02	&	(6.4)		&	(6.4)		&	6.43	$\pm$0.03	\\
{\bfseries $\sigma_\text{Fe}$(keV)}	&	0.11	$\pm$0.04	&	0.07	$\pm$0.06	&	0.36	$\pm$0.10	&	4.82	$\pm$4.79	&	0.15	$\pm$0.06	\\
{\bfseries $\text{E}_\text{gabs 1}$ (keV)}	&	21.76	$\pm$0.06	&	21.37	$\pm$0.10	&	21.82	$\pm$0.10	&	22.72	$\pm$0.60	&	--	--	\\
{\bfseries $\sigma_\text{gabs 1}$(keV)}	&	2.49	$\pm$0.07	&	3.69	$\pm$0.19	&	2.48	$\pm$0.11	&	3.44	$\pm$0.83	&	--	--	\\
{\bfseries $\text{Strength}_\text{gabs 1}$} &	2.95	$\pm$0.10	&	6.53	$\pm$0.64	&	3.41	$\pm$0.17	&	4.08	$\pm$2.15	&	-- --	\\
{\bfseries $\text{E}_\text{gabs 2}$ (keV)}	&	8.38	$\pm$0.14	&	9.11	$\pm$0.16	&	7.76	$\pm$0.59	&	--	--	&	--	--	\\
{\bfseries $\sigma_\text{gabs 2}$(keV)} &	1.19	$\pm$0.19	&	1.89	$\pm$0.33	&	1.18	$\pm$0.52	&	--	--	&	--	--	\\
{\bfseries $\text{Strength}_\text{gabs 2}$}	&	0.12	$\pm$0.02	&	0.43	$\pm$0.16	&	2.18	$\pm$0.10	&	--	--	&	--	--	\\
\hline																
{\bfseries $\chi_{\upsilon}^{2}$ \;/\;d.o.f}	&	383.01	\;/\;324	&	308.36	\;/\;303	&	339.71	\;/\;293	&	228.45	\;/\;232	&	223.4	\;/\;204	\\
{\bfseries Reduced $\chi_{\upsilon}^{2}$} 	&	1.18		&	1.02		&	1.16		&	0.98		&	1.09		\\

\hline										
									
\end{tabular}}
\caption {The best fit spectral parameters of 4U 1538-522 corresponding to NuSTAR Obs II, Obs III, and Obs IV (Pre-eclipse, Ingress and Eclipse). Ingress and Eclipse spectra are fitted with Partial-covering absorber. Photon Index, Cutoff energy ($\text{E}_\text{cut}$), and E-folding energy ($\text{E}_\text{fold}$) are the spectral parameters of \textsc{highecut} model. Flux for each is estimated in (3-79) keV energy range. $\chi_{\upsilon}^{2}$ is the reduced $\chi^{2}$ for the spectral fit. Errors quoted for each parameter are within 1$\sigma$ confidence interval. \label{2}}
\end{center}
\end{table*}

 \begin{figure}
\begin{center}
\includegraphics[angle=270,scale=0.34]{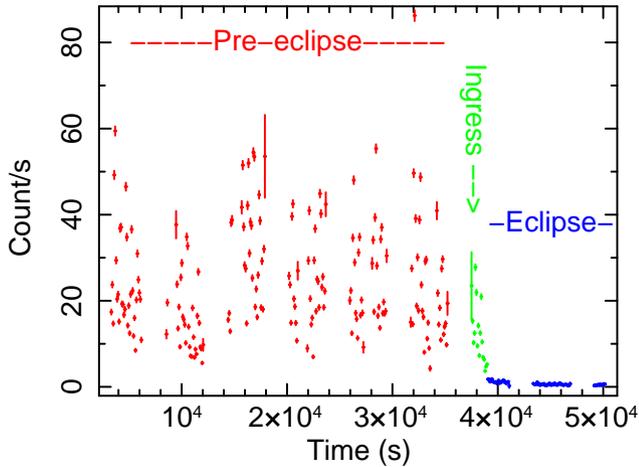}
\caption{The light curve (with start time $\sim 59267.03$ MJD) of the source using NuSTAR observation IV (30602024004) indicating the presence of pre-eclipse, ingress, and an eclipse in red, green, and blue respectively. \label{7}}
\end{center}
\end{figure}

 \begin{figure}
\begin{center}
\includegraphics[angle=0,scale=0.34]{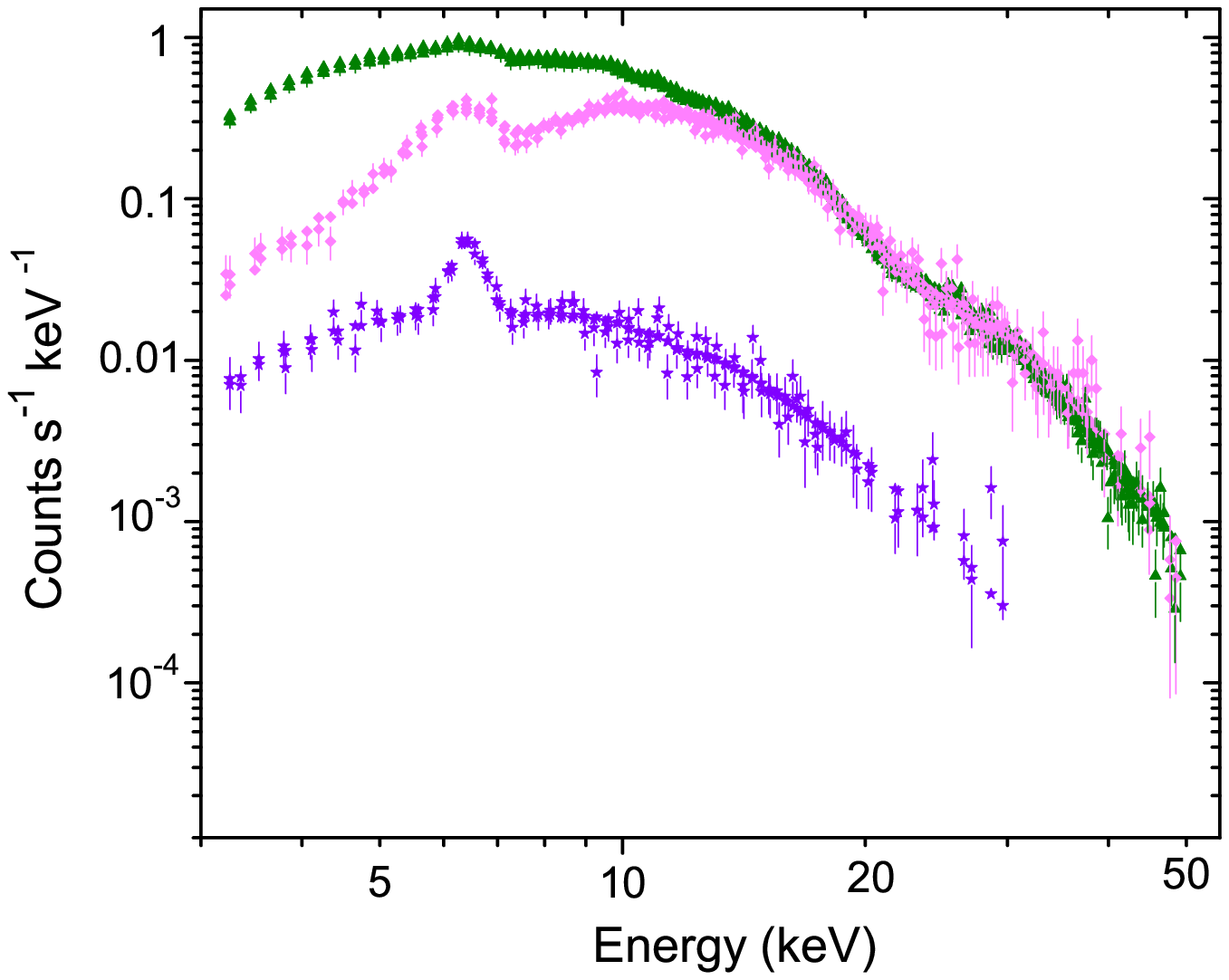}
\caption{NuSTAR spectra in (3-50) keV energy range for the pre-eclipse, ingress, and eclipse corresponding to Obs IV presented in green, magenta, and purple respectively. \textsc{fpma} and \textsc{fpmb} spectra have been combined and rebinned for representative purpose. \label{8}}
\end{center}
\end{figure}

\begin{figure*}
\begin{center}
\includegraphics[width=0.9\linewidth, height=0.5\textheight, angle=0]{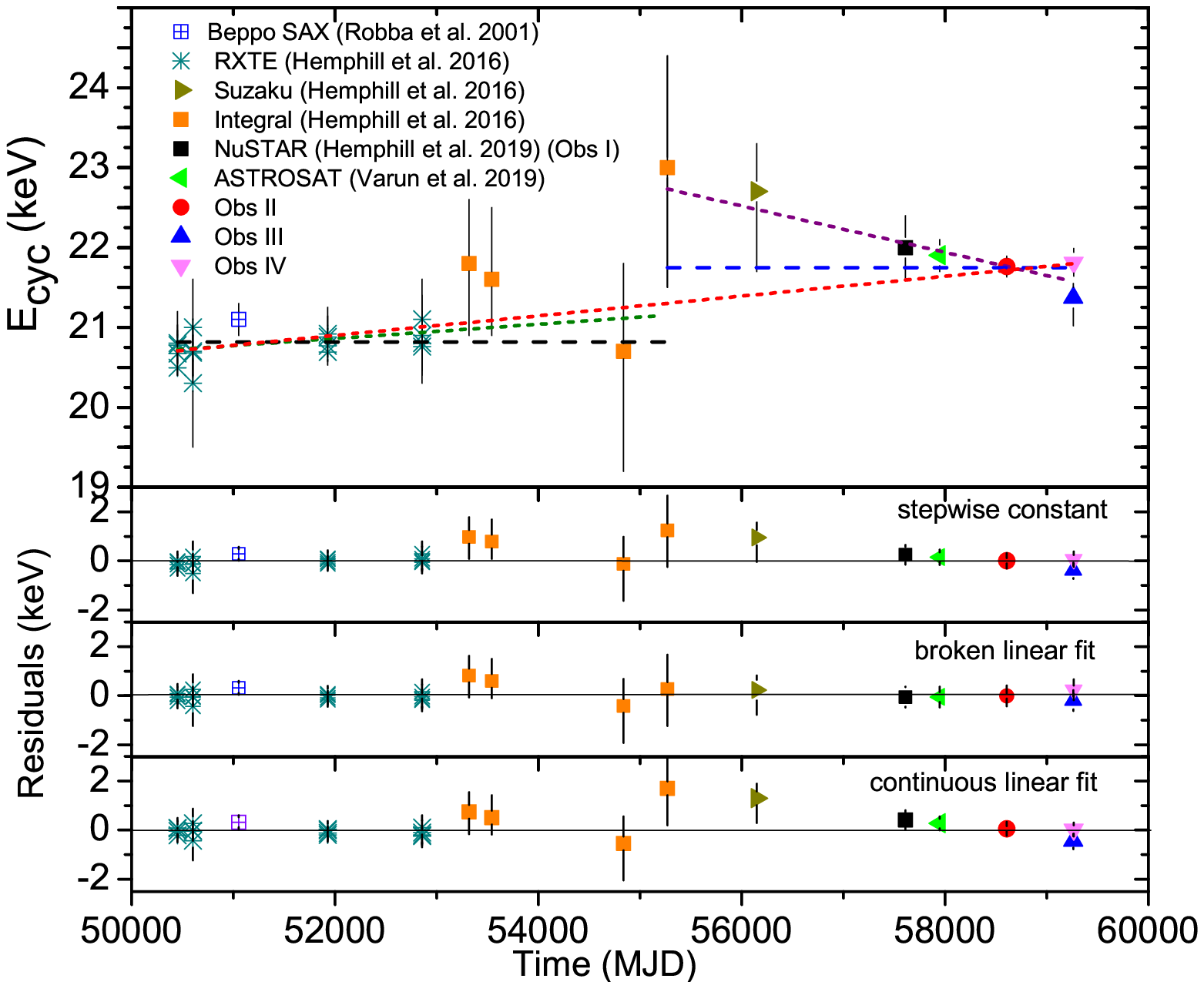}
\caption{Variation of fundamental cyclotron line energy with time as measured with \textit{BeppoSAX} (purple square) by \protect\cite{22}; \textit{RXTE} (pine green asterisk), \textit{Suzaku} (olive green right triangle), and \textit{INTEGRAL} (filled orange square) by \protect\cite{24}; \textit{ASTROSAT} (green left triangle) by \protect\cite{18}; and \textit{NuSTAR} (black square) by \protect\cite{60} along with that obtained by us using the recent NuSTAR observations. The dotted lines in olive green and purple correspond to piecewise sloped fit between (50452.16-54838.25) MJD and (55270.8-59267) MJD, respectively. The red dotted line represents a single continuous fit between (50452.16-59267) MJD. The horizontal dotted lines denote stepwise constant fittings before (black) and after (blue) 55270.8 MJD, respectively. The residuals corresponding to each fitted solutions are shown in the lower three panels. \label{9}}
\end{center}
\end{figure*}

\begin{figure}
\begin{center}
\includegraphics[angle=0,scale=0.32]{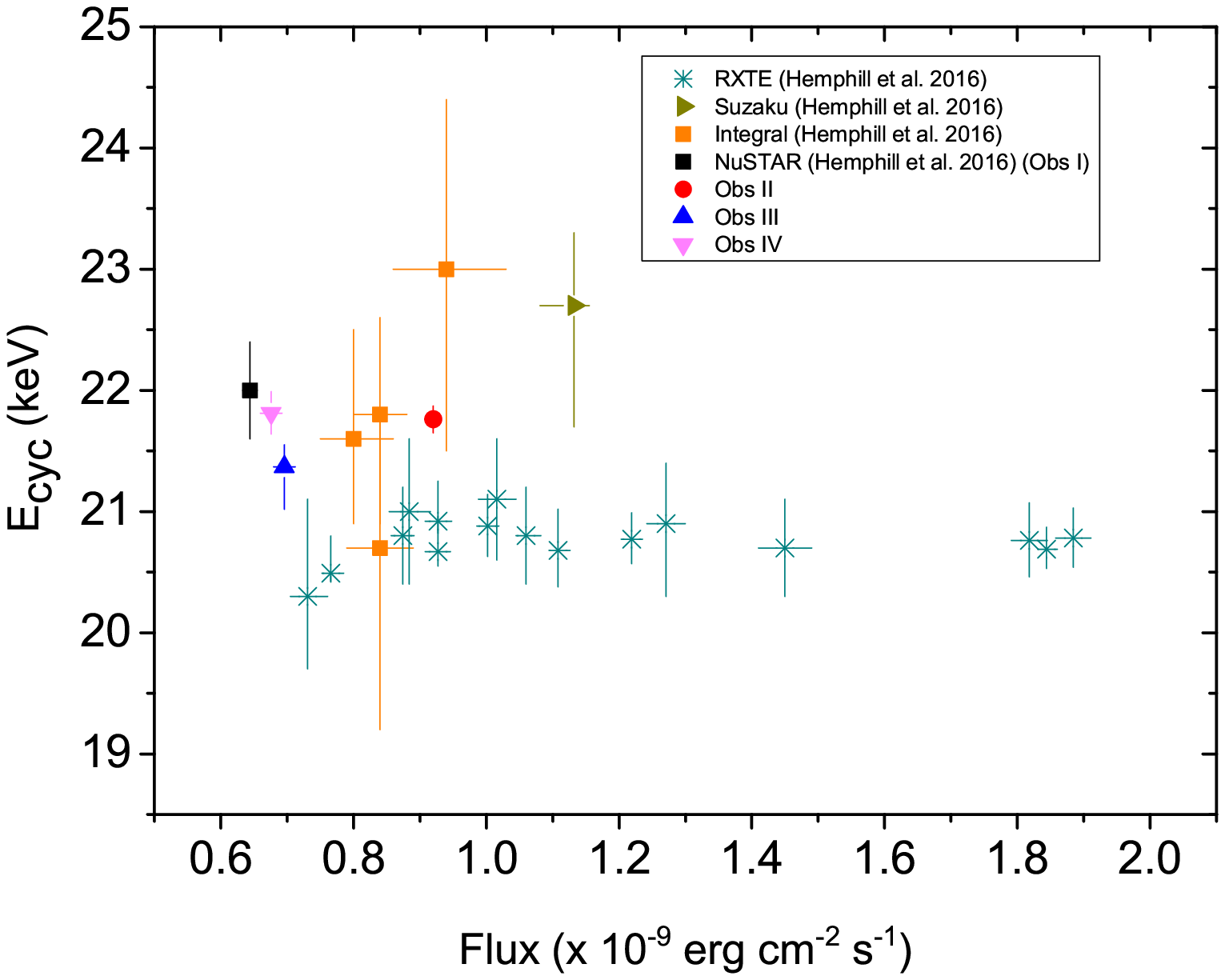}
\caption {Variation of fundamental cyclotron line energy with (3-50) keV flux as measured with \textit{RXTE} (pine green asterisk), \textit{Suzaku} (olive green right triangle), and \textit{INTEGRAL} (filled orange square) by \protect\cite{24}; and \textit{NuSTAR} (black square) by \protect\cite{60} along with that obtained by us using the recent NuSTAR observations. \label{11}}

\end{center}
\end{figure}

\section{Spectral changes in Eclipse}
  The NuSTAR Obs I (30401028002) and IV (30602024004) covers the X-ray eclipse of the source during (57611.81-57612.79) MJD and (59267.04-59267.58) MJD respectively. The source eclipse covered by Obs I is discussed in \cite{60}. Therefore, we study the eclipse covered by recent NuSTAR Obs IV. It covers the orbital phases between (0.776-0.921). Thus, the end of this observation is an ingress along with part of an eclipse. The extracted light curve has been studied in three segments- a pre-eclipse, ingress and a partial eclipse phase which is shown in Figure \ref{7}. We extracted the light curves for each of the phases separately by creating the good time interval (\textsc{gti}) files using tool \textsc{xselect}. Using the \textsc{gti} files, the corresponding spectra for each phase are generated with the help of \textsc{nuproducts}. We fit both \textsc{fpma} \& \textsc{fpmb} spectra for all three phases using the single continuum model \textsc{constant * phabs * tbpcf (highecut* powerlaw + gaussian)* gabs}. We have incorporated partial-covering absorber for fitting both ingress and eclipse spectra. The spectral parameters are presented in Table \ref{2}. We fitted the spectra in (3-50) keV energy range due to low count rate above 50 keV. The pre-eclipse, ingress, and eclipse spectra are presented in Figure \ref{8}. The eclipse spectra exhibits a broad and more prominent emission feature between (6-7) keV. The average flux estimated in (3-79) keV energy range is found as $(4.93 \pm 0.03) \times 10^{-10}  \text{erg\;cm}^{-2}\;\text{s}^{-1}$ whereas the pre-eclipse flux in the same energy range is $(6.91 \pm 0.04) \times 10^{-10}  \text{erg\;cm}^{-2}\;\text{s}^{-1}$.

\begin{figure*}
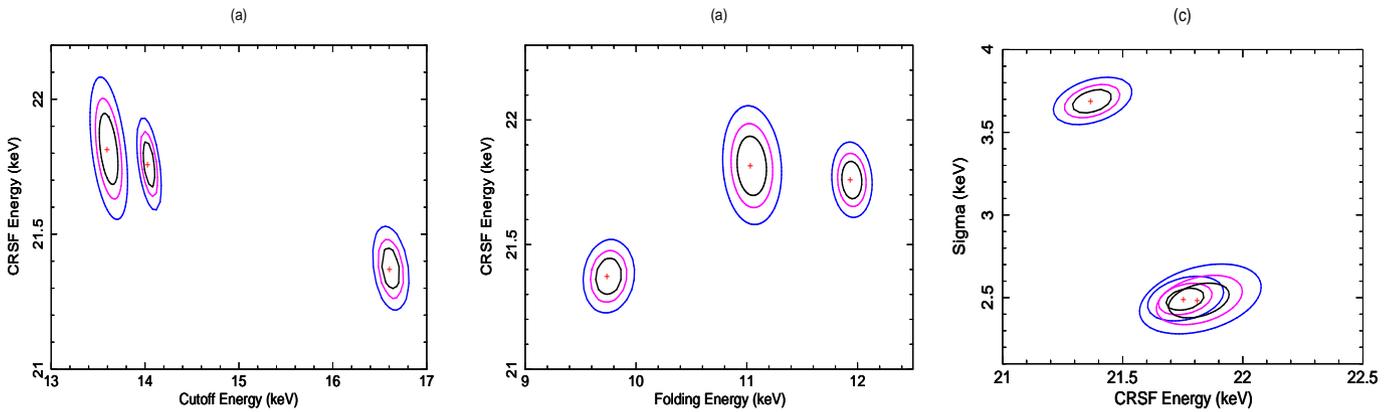

\begin{minipage}{0.3\textwidth}
\includegraphics[width=1.0\linewidth, height=0.25\textheight, angle=270]{ecycecut}
\end{minipage}
\hspace{0.04\linewidth}
\begin{minipage}{0.3\textwidth}
\includegraphics[width=1.0\linewidth, height=0.25\textheight, angle=270]{efecyc}
\end{minipage}
\hspace{0.04\linewidth}
\begin{minipage}{0.3\textwidth}
\includegraphics[width=1.0\linewidth, height=0.25\textheight, angle=270]{sigmaecyc}
\end{minipage}
\caption{(a) $\text{E}_\text{cyc}$ vs. $\text{E}_\text{cut}$, (b) $\text{E}_\text{cyc}$ vs. $\text{E}_\text{fold}$, (c) Cycl. Line width ($\sigma_\text{cyc}$) vs. $\text{E}_\text{cyc}$ contours for NuSTAR observations. Plus sign indicates the best-fit values for each set of contours. The contour (innermost to outermost) corresponds to 68 \%, 90 \%, and 99\% confidence limits. \label{12}}
\end{figure*}

\begin{figure}
\begin{center}
\includegraphics[angle=270,scale=0.3]{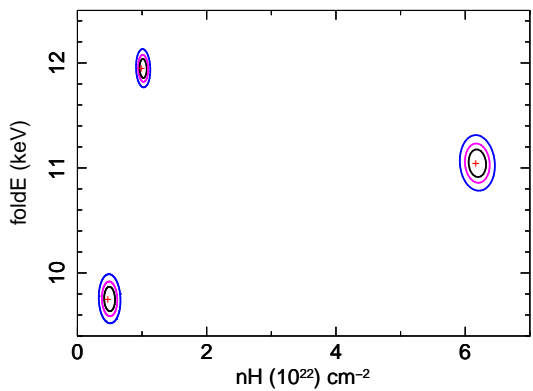}
\caption{Photon Index vs. $E_\text{fold}$ contour for NuSTAR observations. The best-fit value is plotted with plus sign. The contour (innermost to outermost) corresponds to 68 \%, 90 \%, and 99\% confidence limits. \label{13}}
\end{center}
\end{figure}

\section{Discussion \& Conclusion}

\subsection{Pulse periods, Torque reversal \& pulse profiles}

In the paper, we present the spectral and timing analysis of the source 4U 1538-522 considering four NuSTAR observations. The timing analysis of the NuSTAR data detected the pulse period of the source as $ 526.5638 \pm 0.0052$ s, $ 526.7638 \pm 0.0011$ s, $ 526.3135 \pm 0.0032$ s, and $ 526.2341 \pm 0.0041$ s for Obs I, Obs II, Obs III, and Obs IV respectively.

The pulse period measurements of the source indicate the presence of torque reversal $\sim$ 58620 MJD which is in agreement with the  \textit{Fermi}/GBM data. The previous studies of the source reported torque reversal in late 2008 or early 2009 by \cite{4} with the spinning down of the source at the rate $\sim 0.13$ s $\text{yr}^{-1}$. We roughly estimated the spin down rate again including more \textit{Fermi}/GBM data. The source is found to exhibit spin down trend at the rate  $\dot{P} $= 0.163 $\pm$ 0.002 s $\text{yr}^{-1}$ until $\sim$ 58603 MJD. \textsc{batse} monitoring between 1991 to 2000 reveals a spinning up of the source at the rate $\dot{P} $= -0.255 $\pm$ 0.004 s $\text{yr}^{-1}$. At time $\sim$ 58620 MJD, we found an evidence of torque reversal after which there is a strong spinning up of the source at the rate roughly estimated as $\dot{P} $= -0.305 $\pm$ 0.018 s $\text{yr}^{-1}$ until 59275 MJD which is again followed by a decreasing and increasing trend according to \textit{Fermi}/GBM data.  The change in accretion rate leads to a change in pulse period of the source. As the Neutron Star (NS) rotates, its magnetic field lines rotate along with it and the accreting matter exert a torque onto the NS. The relationship between accretion rate and pulse period is easier to understand whereas that between accretion rate and torque is very complex. 4U 1538-522 being a wind fed system, the infalling material may or may not impart sufficient angular momentum so as to form a persistent accretion disk \citep{4}. A torque reversal may arise due to a significant change in interaction region between the infalling matter and the magnetic field of the NS. The torque reversal is typically related with a significant variation in luminosity and/or a dramatic spectral shifts \citep{b}. In 4U 1538-522, such significant change in luminosity is not observed before and after torque reversal. We observed a marginal insignificant temporal variation in the pulse profile before and after reversal (Figure \ref{2}). The rate of spinning up of the source after the torque reversal is found to be only $\sim$ 2 times the rate of its spinning down before the torque reversal. In the case of 4U 1907+09, \cite{d} explained such torque reversals without large changes in luminosity with model of \cite{f}. The model requires an Alfvén radius comparable to the corotation radius of the neutron star to produce the localized propeller effect. According to a standard theory, for an accreting neutron star, the accreted material reaches the NS surface only if the radius of the magnetosphere of the neutron star is lesser than the corotation radius \citep{a}. However, if the magnetospheric radius is greater than or equal to the corotation radius, the matter is thrown out beyond the capture radius by the centrifugal barrier generated by the magnetic field of the NS. The neutron star then enters the propeller regime which results in a distinct evolution in its pulse period. For 4U 1538-522, its corotation radius is (1-2) orders of magnitude larger than its Alfvén radius if spherical accretion is assumed \citep{c}.  \cite{3} suggested that the only possible reason behind the spin-up trend of 4U 1538-522 observed between 1991-1995 was a consequence of a random walk in pulse frequency derivative which originated from a random walk in accretion torque on the NS. The spin up-trend followed after torque reversal between (2019-2021) may not be just due to random walking of the system. Thus, both of these theories seems to be insufficient to explain the torque reversal observed. Such cases in which the pulsars are found in spin states with opposite sign but nearly equal magnitudes is practically impossible to explain within the framework of standard accretion torque theory \citep{3}. Thus, the detailed analysis of the observed torque reversal would have been possible if there were more number of observations around the torque reversal. For general discussion on different accretion models and their predictions for torque changes we refer \cite{mmmm}. A comprehensive study associated with the magnetospheric radii, propeller effect, etc., has been carried out by \cite{bbbb}, emphasizing on disc-magnetosphere interaction models in lower luminosity accreting systems.

\begin{table*}
\begin{center}
\begin{tabular}{l|l|l|l|l}
\hline											
											
\bfseries Fit Solutions &	\bfseries Number of Points &	\bfseries Degrees of Freedom	&	\bfseries Reduced  $\chi^{2}$	&	\bfseries Pearson's r	\\
\hline
\bfseries Single continuous	&	27	&	25	&	0.9	&	0.88	\\
\hline
\multirow{2}{*}{\bfseries Stepwise constant}	&	21	&	20	&	0.51	&	0.35	\\
	&	7	&	6	&	1.47	&	-0.79	\\
\hline
\multirow{2}{*}{\bfseries Broken linear}	&	21	&	19	&	0.48	&	0.35	\\
	&	7	&	5	&	0.67	&	-0.79	\\
\hline										
\end{tabular}
\caption {Fit parameters and  $\chi_{\upsilon}^{2}$ obtained corresponding to single-continuous, stepwise constant, and broken linear fit solutions. \label{3}}
\end{center}
\end{table*}

The evolution of pulse profile with energy is consistent with the results obtained from Ginga \citep{21}, BeppoSAX \citep{22}, Suzaku \citep{23}, and \textsc{astrosat} \citep{18} for the source 4U 1538-522. The pulse profile resembles a double-peaked nature with primary and secondary peak. We observe a similar variation in the pulse profile with energy relative to the previously reported results. The secondary peak is found to become weaker near cyclotron line energy (energy band 18-24 keV) as can be seen in Figure \ref{1}. A change in shape of the pulse profile or pulse fraction near the cyclotron line energy has been seen in several pulsars like 1A 1118–61, A 0535+26, XTE J1946+274 \citep{65} and V 0332+53 \citep{66}. Based on the probes attempted by \cite{66}, \cite{70}, \cite{71}, there is a significant alteration and an increase by a large factor of the scattering cross sections near the CRSF which causes the change in the shape of the pulse profile near the corresponding resonance energy. It is noted that a non-monotonic dependence of the pulse fraction exist with energy  which is typical for many pulsars with cyclotron lines as reported earlier \citep {pf, 70}. The change in scattering cross-section near the cyclotron line energy affects the beaming pattern and emission geometry which causes a variation in the pulse intensity \citep{011, 9}. Hence, there is a change in the pulse fraction as it provides a quantitative description of the pulse intensity \citep{W22}. The resonant scattering at the Landau level causes a formation of the local minima in the P.F vs. energy plot due to a significant decrease in the photon count around $\sim$ 21 keV \citep{pf}.

The average flux in the energy range (3-79) keV corresponding to the NuSTAR Obs III is estimated as $ (7.05 \pm 0.02) \times 10^{-10}  \text{erg\;cm}^{-2}\;\text{s}^{-1}$ with the corresponding luminosity $\sim 3.92 \times 10^{36} \text{erg\;s}^{-1}$ assuming a distance of $\sim$ 6.82 kpc. For Obs IV, the average flux estimated in (3-79) keV energy range is found to be $(4.93 \pm 0.03) \times 10^{-10}  \text{erg\;cm}^{-2}\;\text{s}^{-1}$ whereas the pre-eclipse flux in the same energy range is $(6.91 \pm 0.04) \times 10^{-10}  \text{erg\;cm}^{-2}\;\text{s}^{-1}$. The source 4U 1538-522 belongs to  a category of sources that shows a peculiar feature around $\sim$ 10 keV. Therefore, we incorporated \textsc{gabs (smooth)} component to the continuum model while fitting the spectrum. The feature observed $\sim$ 10 keV is typically known as ‘10 keV bump’ or ‘10 keV feature’ and is found in many pulsars such as 4U 1907+09, V0331+53, EXO 2030+375 etc. \citep{20001}. However, the cause of occurence is yet to be understood clearly. Although in some cases the 10 keV-feature is interpreted as a separate physical
component but most probably it reflects the drawback of the simple phenomenological models used in spectral fitting.
To study the acceptability of the CRSF parameters, we plotted the $\chi^{2}$ contours between pair of parameters corresponding to Obs II, III, IV using \textsc{xspec} as shown in Figure \ref{12}. The $E_\text{cyc}$ vs. $E_\text{cut}$ and $E_\text{cyc}$ vs. $E_\text{fold}$ contours shows a weak or no correlation for the continuum model under consideration. Thus, the contour plots show that the possible variations in centroid energy of the fundamental cyclotron line are not due to variation in model parameters. There is an intrinsic correlation between cyclotron line width ($\sigma_\text{cyc}$) and cyclotron line energy ($E_\text{cyc}$) as shown in Figure \ref{12}. No degeneracy is observed between Photon index $\Gamma$ and E-folding energy ($E_\text{fold}$) as presented in Figure \ref{13} corresponding to Obs II, III, and IV.

\subsection{Variation of \textbf{$E_\text{cyc}$} with time and luminosity}

According to \cite{24}, previous studies of the variation in CRSF energy reported a long term increase in its centroid energy by $\sim 1.5$ keV between 1996 -2004 (RXTE) and 2012 Suzaku measurements. They reported a linear increasing rate of 0.058 $\pm$ 0.014 keV $\text{yr}^{-1}$. ASTROSAT measurements by \cite{18} were also consistent with results obtained by \cite{24}. The CRSF centroid energy for the NuSTAR Obs I has been previously reported to lie at an energy $\sim$ 22 keV by \cite{60}. For Obs II, III, and IV (pre-eclipse), we determine the CRSF centroid energy as $\sim$ 21.76 keV, $\sim$21.37 keV, and  $\sim$21.82 keV respectively. In order to understand the temporal variation of $E_\text{cyc}$, we incorporated our measurements from NuSTAR observations to the earlier  $E_\text{cyc}$ measurements (refer Figure 11 from \cite{60}) as shown in Figure \ref{9}. The CRSF centroid energy was found to be apparently constant for the previous measurements carried out with RXTE and BeppoSAX observations. However, the line energy measurements revealed higher values that are estimated using NuSTAR and ASTROSAT observations. Interestingly, the INTEGRAL and Suzaku observations reported by  \cite{24} reveals a change in the trend of the cyclotron line energy $\sim$ 55270.8 MJD, but the large uncertainties associated with the energy measurements detains us from making any firm statement.

We performed fitting of the temporal variation of $E_\text{cyc}$ measurements and found three basic solutions: (i) a linear fit with a rising slope and no discontinuity between (50452.16-59267) MJD; (ii) a "broken" linear fit with a rising and falling slope between $\sim$ (50452.16-55270.8) MJD and (55270.8-59267) MJD, respectively, with a jump in $E_\text{cyc}$ between the latter two INTEGRAL measurements; and (iii) a "stepwise" constant solution, again with a jump in $E_\text{cyc}$ as presented in Figure \ref{9}. The lower panels in Figure \ref{9} represent residuals corresponding to each fitted solution. The parameters for each fit, along with the obtained fit statistics, are presented in Table \ref{3}. Recent data, including the measurements presented in this paper, indicates a decrease in $E_\text{cyc}$ after $\sim$ 55270.8 MJD. However, the $E_\text{cyc}$ measurements after $\sim$ 55270.8 MJD lie within the uncertainties in the Integral measurements, as presented by \cite{24}. A linear fit to the entire data yields an increase in $E_\text{cyc}$ at 0.05 $\pm$ 0.01 keV $\text{yr}^{-1}$. Broken linear fit up to $\sim$ 55270.8 MJD shows an increase in $E_\text{cyc}$ at 0.03 $\pm$ 0.02 keV $\text{yr}^{-1}$ followed by a decrease at -0.11 $\pm$ 0.04 keV $\text{yr}^{-1}$. The step-wise constant fitted solution of $E_\text{cyc}$ data indicates a stable variation around $\sim$ 20.81 $\pm$ 0.05 keV and $\sim$ 21.75 $\pm$ 0.09  keV, before and after $\sim$ 55270.8 MJD, respectively. Pearson's correlation coefficient (r) corresponding to both stepwise-constant and broken-linear are found to be $\sim$ 0.35  and -0.79, before and after $\sim$ 55270.8 MJD, respectively. This reveals the existence of a low positive correlation of $E_\text{cyc}$ before $\sim$ 55270.8 MJD, followed by a high anti-correlation. The overall linear fitting shows a significant temporal correlation of $E_\text{cyc}$ (r= 0.88).  The reduced $\chi^{2}$ obtained for each fit suggests that the temporal variation of $E_\text{cyc}$ measurements before $\sim$ 55270.8 MJD is better explained by the constant fitted solution than the broken linear one. However, the broken linear fitted solution explains well the variation of $E_\text{cyc}$ measurements with time after $\sim$ 55270.8 MJD. We performed an F-test assuming the constant fit solution as the null hypothesis to investigate the statistically better model between the two. The F-test probability for the $E_\text{cyc}$ variation before and after $\sim$ 55270.8 MJD, is found to be 0.12 and 0.03, respectively. For the overall variation between (50452.16-59267) MJD, we found the F-test probability as < 0.1 $\times$ $10^{-4}$, indicating the signifance of linear fit over the constant one. Thus, the F-test strengthens the result that the step-wise constant fitted solution is statistically better than the broken-linear fit for $E_\text{cyc}$ variation with time before  $\sim$ 55270.8 MJD. However, after $\sim$ 55270.8 MJD, the linear fitted solution is statistically better than the constant fitted solution. In all the three cases, the reported results are above 3 $\sigma$ confidence level.

A significant "jump" in CRSF energy was reported and discussed earlier by \cite{24, 60} with an INTEGRAL measurement of $\sim$ 55270.8 MJD. As discussed by \cite{24}, the possible reason behind the sudden increase in CRSF energy could be the reconfiguration of the mound geometry. The geometric displacement of the emission region or a change in the configuration of the local field results in the shift in $E_\text{cyc}$ \citep{61}. The variation in $E_\text{cyc}$ may be due to an imbalance between the inflow and outflow of the material, which causes a change in the accretion mound with time \citep{27}. If the inflow slightly exceeds the outflow, there is a slow accumulation in the mound, causing an increase in its height or a change in the local field structure, leading to a variation in $E_\text{cyc}$. When there is an excess accumulation, the pressure in the accretion mound becomes too high, and the outflowing material is forcefully ejected, resulting in the adjustment of the accretion mound to its original form \citep{555}. The event is relatively abrupt and violent, which possibly explains the sudden upward jump in CRSF energy observed $\sim$ 55270.8 MJD. However, if the outflow slightly exceeds the inflow, the slow accumulation of matter possibly increases the mound height, resulting in a change in the local field structure, as explained by \cite{555}. This might be the reason for the decrease in CRSF energy observed after $\sim$ 55270.8 MJD. As suggested by \cite{18}, the long- term evolution in $E_\text{cyc}$ may be ascertained by more observations in the future that are separated by longer time intervals.

 In case of an accreting neutron star, there are generally two regimes associated with the rate of accretion i.e., sub-critical (low-luminosity) and super-critical (high-luminosity) regimes \citep{1326}. The transition between these two regimes occurs at the critical luminosity ($\text{L}_{\text{c}}$).  At low luminosity, the infalling material forms an accretion mound in the polar cap and the radiation escapes from the mound along magnetic field lines \citep{6021, 5021} resulting in 'pencil-beam' emission. At high luminosity, the radiation pressure decelerates the infalling material and a radiation-dominated shock is formed in the polar cap. The radiation mainly escapes through the walls of the accretion column resulting in 'fan-beam' emission \citep{1326, 1221}. The variations of cyclotron line energy with luminosity has been found in a number of X-ray transients such as V 0332+53 \citep{66,75}, A 0535+26 \citep{85}, Vela X-1 \citep{87, 89}, GX 304-1 \citep{90, 91}, Cep X-4 \citep{92}, and 4U 0115+63 \citep{93, 94, 95}. The sources of relatively low mass accretion rates are found to exhibit a positive correlation between the cyclotron line energy and luminosity. The negative correlation between $\text{E}_\text{cyc}$ and luminosity is observed in the sources with high mass accretion rates. \cite{96} reported both positive as well as negative correlations in the case of V0332+53. The positive correlation was detected at low luminosities and the negative correlations at a high luminosity state. The relation between the CRSF centroid energy and accretion luminosity does not follow a specific trend as the same source may exhibit both positive and negative correlations between $\text{E}_\text{cyc}$ and luminosity depending upon their accretion state \citep{97}. 
   To study the variation of the CRSF centroid energy with flux, we incorporated results from our work with the earlier findings by \cite{24, 60} as shown in Figure \ref{11}. As discussed in \cite{24}, the source 4U 1538-522 is reasonably predicted to be in super-critical state. The range of critical luminosity is $\text{L}_{\text{c}} \sim (2-4)\times10^{36} \text{erg\;s}^{-1}$ for wind-accreting sources and is comparatively higher than for disc-accreting sources \citep{411}. The unabsorbed 3-50 keV flux for Obs II, III and IV (pre-eclipse) are in the range of $\sim$ (6.76-9.16) $\times 10^{-10} \text{erg\;cm}^{-2}\;\text{s}^{-1}$, which corresponds to a luminosity in the band (3.76-5.09) $\times10^{36} \text{erg\;s}^{-1}$ at a distance of $\sim$ 6.82 kpc and is comparable to the low-flux band of RXTE observations \citep{24}. Thus, our measurements are consistent with \cite{24} and strengthens the placement of the source in super-critical regime as its critical luminosity is $\sim$ $10^{36} \text{erg\;s}^{-1}$ \citep{24}.

\subsection{An X-ray Eclipse}

 The NuSTAR Obs IV (30602024004) partially covers the X-ray eclipse of the source at an expected orbital phase (Table \ref{1}). The eclipse covered by the observation has a duration of $\sim$ 10869 seconds. However, the complete X-ray eclipse and the egress phase is not covered by this observation with 21.6 ks exposure. The mid-eclipse time of the source cannot be determined precisely due to the limitation of full coverage of the eclipse. We divided the light curve into three segments; pre-eclipse, ingress and an eclipse phase as shown in Figure \ref{7}. In order to study the variation of flux and corresponding luminosities across different phases, we fitted the spectra for each of them individually. The eclipse flux is found to drop by $\sim$ 96.96\% which is similar to flux drop reported by \cite{g} in \textit{Chandra}-HETGS observations and \cite{60} in NuSTAR observation. The partial eclipse covered by Obs IV has a duration of $\sim$ 0.126 d which is shorter than the NuSTAR eclipse reported by \cite{60} with Obs I. The partial-covering absorption model is chosen to fit the ingress and eclipse spectra. During an eclipse, the spectral changes is assumed to be due to absorbing and scattering of material in the line of sight. As a result of the change in absorption, we found an increase in both $\text{n}_{\text{H},\text{c}}$ and partial covering fraction ($\text{f}_{\text{p},\text{c}}$) during ingress while fitting the spectrum with a partial-covering absorber consistent with \cite{60}. This is expected if QV Nor's wind is composed of small, dense clumps. These clumps shadows the light of sight when the pulsar enters into the limb of QV Nor \citep{60}.
 
  The pre-eclipse flux is comparatively larger than the flux measured from the average spectra in the energy range of (3-79) keV. The CRSF line energy detected in the ingress spectra is found to be slightly higher than in the pre-eclipse and eclipse spectra consistent with \cite{60}. During the X-ray eclipse, the neutron star and the X-ray source are behind the massive companion. The direct X-ray radiation coming from the compact object is blocked by the companion. Thus, the X-ray radiations observed during the eclipse are the reprocessed emission as a result of the interaction of primary X-rays with the surrounding medium i.e., the stellar wind that acts as the main reprocessing agent in the case of HMXBs. The emission lines are less suppressed during the eclipse due to which the spectrum shows emission lines with larger equivalent width \citep{222}. Hence, we observed an expected broad emission feature between (6-7) keV in the continuum as shown in Figure \ref{8}.

\section*{ACKNOWLEDGEMENT}
 
 This research has made use of the publicly available data of the pulsar provided by High Energy Astrophysics Science Archive Research Center (HEASARC) Online Service data archive. The \textit{Fermi}/GBM data used in our study is provided by \textit{Fermi}-GAPP team. This work has made use of data from the European Space Agency (ESA) mission
{\it Gaia} (\url{https://www.cosmos.esa.int/gaia}), processed by the {\it Gaia} Data Processing and Analysis Consortium (DPAC, \url{https://www.cosmos.esa.int/web/gaia/dpac/consortium}). Funding for the DPAC has been provided by national institutions, in particular the institutions participating in the {\it Gaia} Multilateral Agreement. RT would like to acknowledge CSIR/NET for a research grant 09/0285(11279)/2021-EMR-I. We would like to thank anonymous reviewer for his/her valuable comments and suggestions which helped us in improving the manuscript both in quality and quantity. Authors are thankful to IUCAA Center for Astronomy Research and Development (ICARD), Physics Dept, NBU for extending research facilities.
 
\section*{DATA AVAILABILITY}
 We have used the publicly available observational data accessed from the HEASARC data archive for carrying out this research work.

%%%%%%%%%%%%%%%%%%%%%%%%%%%%%%%%%%%%%%%%%%%%%%%%%%



%\bibliographystyle{mnras}


%%%%%%%%%%%%%%%%%%%%%%%%%%%%%%%%%%%%%%%%%%%%%%%%%%

%%%%%%%%%%%%%%%%% APPENDICES %%%%%%%%%%%%%%%%%%%%%

%%%%%%%%%%%%%%%%%%%%%%%%%%%%%%%%%%%%%%%%%%%%%%%%%%

% Don't change these lines
\bsp	% typesetting comment
\label{lastpage}
\end{document}